\documentclass[%
 reprint,
 superscriptaddress,
 amsmath,amssymb,
 aps,
 prfluids, 
 onecolumn,
 floatfix
]{revtex4-2}

\usepackage{graphicx}
\usepackage{siunitx}

\usepackage[usenames,dvipsnames]{xcolor}

\newcommand{\hypgeo}[2]{%
  \operatorname{%
    {\vphantom{\mathnormal{F}}}_{#1}%
    \kern-\scriptspace
    \mathnormal{F}_{#2}%
  }%
}

\usepackage[%
    linkcolor = blue, 
    citecolor = blue, 
    urlcolor = blue, 
    colorlinks = true
]{hyperref}

\begin{document}

\title{Flow rate--pressure drop relations for shear-thinning fluids in deformable configurations: theory and experiments}

\author{SungGyu Chun}
\thanks{These authors contributed equally}
\affiliation{Department of Mechanical Science and Engineering, University of Illinois at Urbana-Champaign, Urbana, Illinois 61801, USA}
\author{Evgeniy Boyko}
\thanks{These authors contributed equally}
\affiliation{Faculty of Mechanical Engineering, Technion -- Israel Institute of Technology, Haifa 3200003, Israel}
\author{Ivan C. Christov}
\email[Email address for correspondence: ]{christov@purdue.edu}
\affiliation{School of Mechanical Engineering, Purdue University, West Lafayette, Indiana 47907, USA}
\author{Jie Feng}
\email[Email address for correspondence: ]{jiefeng@illinois.edu}
\affiliation{Department of Mechanical Science and Engineering, University of Illinois at Urbana-Champaign, Urbana, Illinois 61801, USA}
\affiliation{Materials Research Laboratory, University of Illinois at Urbana-Champaign, Urbana, Illinois 61801, USA}

\date{\today}

\begin{abstract}
We provide an experimental framework to measure the flow rate--pressure drop relation for Newtonian and shear-thinning fluids in two common deformable configurations: (\textit{i}) a rectangular channel and (\textit{ii}) an axisymmetric tube. Using the Carreau model to describe the shear-dependent viscosity, we identify the key dimensionless rheological number, $Cu$, which characterizes shear thinning, and we show that our experiments lie within the power-law regime of shear rates. To rationalize the experimental data, we derive the flow rate--pressure drop relation taking into account the two-way-coupled fluid--structure interaction between the flow and its compliant confining boundaries. We thus identify the second key dimensionless number, $\alpha$, which characterizes the compliance of the conduit. We then compare the theoretical flow rate--pressure drop relation to our experimental measurements, finding excellent agreement between the two. We further contrast our results for shear-thinning and Newtonian fluids to highlight the influence of $Cu$ on the flow rate--pressure drop relation. Finally, we delineate four distinct physical regimes of flow and deformation by mapping our experimental flow rate--pressure drop data for Newtonian and shear-thinning fluids into a $Cu-\alpha$ plane.
\end{abstract}

\maketitle 

\section{Introduction}

Elucidating the relationship between the pressure drop, $\Delta p$, and the flow rate, $q$, in different geometries plays a central role in understanding hydrodynamic features across a diverse set of scientific fields ranging from mechanical \cite{ozsun2013non,mehboudi2019experimental} and biomedical \cite{grotberg2004biofluid,bhatia2014microfluidic,parthasarathy2022elastic} engineering, to wearable diagnostics \cite{kashaninejad2023microfluidic}, soft robotics \cite{elbaz2014dynamics,polygerinos2017soft}, and flow control in plants' vasculature \cite{park2021fluid,keiser2022intermittent}, to name a few. However, while the flow rate--pressure drop relations for laminar flow of Newtonian fluids in common geometries are well understood~\cite{happel1983}, this is not the case for non-Newtonian fluids flowing through either rigid~\cite{boyko2021flow} or deformable~\cite{christov2021soft} conduits. Even {as microfluidic techniques for shear viscometry of complex fluids are gaining popularity \cite{pipe2009microfluidic,gupta2016microfluidic} (in particular, to study nonlinear flow rate--pressure drop relationships \cite{suteria2018nicrofluidic}), a complete understanding of how the interplay between shear-thinning rheology and wall compliance sets the flow rate--pressure drop relation for steady low-Reynolds-number flow of a shear-thinning fluid in a deformable configuration is still lacking.} Furthermore, there are no thorough \emph{quantitative} comparisons of the theoretical predictions for the flow rate--pressure drop relation with experimental data. Thus, the twofold aim of our study is to understand the interplay between shear-thinning rheology and wall compliance \emph{via} detailed quantitative comparisons between theory and experiments.

Previous studies of the pressure-driven flow of shear-thinning fluids in deformable configurations at low Reynolds numbers were either solely experimental \cite{delg2016is,raj2018flow,raj2019biomimetic,nahar2019influence} or theoretical \cite{anand2019non,anand2021revisiting}. For example, \citet{raj2016flow} and \citet{delg2016is} measured the pressure drop due to shear-thinning fluid flow through a microchannel with a deformable wall, yet no theory was proposed to capture the experimental observations. 
\citet{raj2018flow,raj2019biomimetic} conducted experiments of non-Newtonian fluid flow in a compliant cylindrical conduit using a shear-thinning xanthan gum solution. However, their experiments exhibited only weak fluid--structure interaction (FSI), and thus deviations from the rigid-tube $q-\Delta p$ relation for a shear-thinning fluid (within the power-law regime) were negligible. 
Motivated by experimental observations, \citet{anand2019non} and \citet{anand2021revisiting} initiated the development of predictive theory for flow-induced deformation of compliant rectangular channels and axisymmetric tubes, respectively. They considered a power-law model for the shear-dependent viscosity to describe shear thinning and validated their theoretical predictions for the nonlinear flow rate--pressure drop relation against simulations. 

Any theoretical prediction in non-Newtonian fluid mechanics inherently relies on a specific constitutive model that aims to describe a certain rheological behavior, such as shear thinning, viscoelasticity, etc. 
However, theoretical predictions do not always agree with the experimental observations, sometimes even qualitatively. In fact, there are many discrepancies between experimental and theoretical results of non-Newtonian fluid flows, for example, the flow rate--pressure drop relation for viscoelastic Boger fluids flowing through rigid contraction and contraction--expansion geometries~\cite{koppol2009anomalous,sanchez2022understanding}. While experiments show a nonlinear increase in the pressure drop with the flow rate~\cite{rothstein1999extensional,rothstein2001axisymmetric}, theory and numerical simulations based on continuum dumbbell models predict a nonlinear decrease in the pressure drop~\cite{alves2003benchmark,binding2006contraction,boyko2022pressure}.
Therefore, given these experiment/theory discrepancies, a quantitative comparison of any theoretical result with experimental data is of fundamental importance since it provides insight into the adequacy of the constitutive model used. However, despite the importance of quantitative assessments, no prior study has made this comparison for the flow rate--pressure drop relation of shear-thinning fluids in deformable configurations. One possible reason is that previous experimental works did not have a complete theory, identifying the key dimensionless parameters governing this multiphysics problem, to guide a systematic experimental investigation. 

In this work, we combine experiments and theory to elucidate the interplay between shear-thinning rheology and wall compliance on the flow rate--pressure drop relation of deformable configurations and to enable a quantitative comparison between the theoretically-predicted $q-\Delta p$ curves and the experimental data. For two common configurations, (\textit{i}) a rectangular channel and (\textit{ii}) an axisymmetric tube, we identify two key dimensionless numbers, the Carreau number $Cu$ and the compliance number $\alpha$, which characterize shear thinning and compliance of the conduit, respectively, and utilize them to delineate four distinct physical regimes of flow and deformation of the $q-\Delta p$ data.

\section{Problem formulation}

\begin{figure}
    \centerline{\includegraphics[scale=1.2]{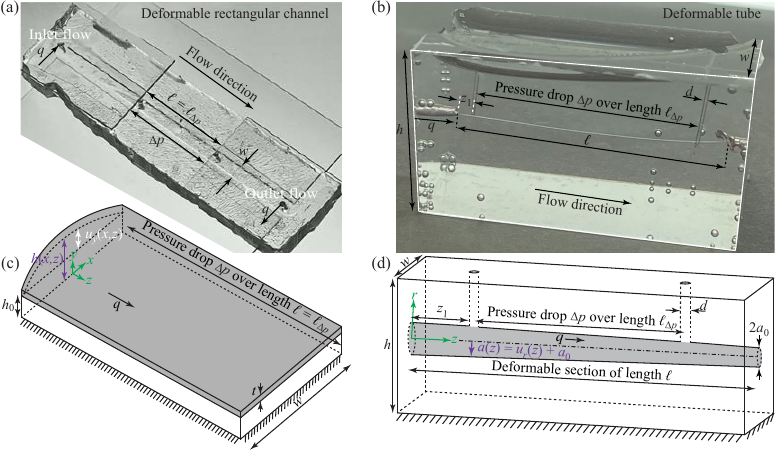}}
    \caption{Illustration of the configurations used for experiments and modeling. (a,c) Image of the (a) experimental device and (c) schematic of a 3D channel of an initially rectangular cross-section with a top deformable wall. (b,d) Image of the (b) experimental device and (d) schematic of a 3D tube extruded from a large block of elastic material, exhibiting radial deformation.  
    Both configurations have a deformable section of length $\ell$ and contain either a Newtonian or shear-thinning fluid steadily driven by the imposed flow rate $q$. Our interest is to determine the pressure drop $\Delta p$ over a streamwise distance $\ell_{\Delta p}$ between two pressure ports. For the channel, we have $\ell \approx \ell_{\Delta p}$, while for the tube $\ell > \ell_{\Delta p}$.}
    \label{F1}
\end{figure}

We study the steady fluid--structure interaction between incompressible Newtonian and shear-thinning fluids and two different complaint geometries, as shown in Fig.~\ref{F1}. We consider (a,c) a 3D channel of initially rectangular cross-section with a top deformable wall and (b,d) a 3D cylindrical exclusion within a large rectangular slab, exhibiting axisymmetric radial deformation. The configurations have a compliant section of length $\ell$. Two pressure ports are introduced, as shown in Fig.~\ref{F1}, measuring the pressure drop $\Delta p$ over a streamwise distance $\ell_{\Delta p}$. The rectangular channel has width $w$ and height $h$, where $h\ll w\ll\ell$; the undeformed height is $h_0$. The tube has a radius $a$; the undeformed radius is $a_0$. We impose the volumetric flow rate $q$ at the inlet, which produces a flow with velocity field $\boldsymbol{v}$ and pressure distribution $p$. In this work, we characterize experimentally the steady-state relation between the pressure drop $\Delta p$ and the flow rate $q$ for Newtonian and shear-thinning fluids in two configurations and then perform a quantitative comparison between the experimental results and theoretical predictions.

\section{Experimental set-up and material characterization}

\subsection{Design and fabrication of the deformable configurations}

For the rectangular channel with a deformable top wall, we first used a 3D printing technique (ELEGOO Mars Resin 3D Printer) to fabricate the mold with the designed channel dimensions. We prepared the polydimethylsiloxane elastomer (PDMS, Sylgard 184, Dow Corning, MI, USA) with a mixing ratio of 10:1 (w/w) between the silicone elastomer base and the curing agent. Then we poured the PDMS mixture onto the 3D printed mold and degassed the mixture under vacuum for an hour to remove excess air bubbles. We cured the mixture in an oven at $90\si{\degreeCelsius}$ for 12 hours. After curing, we carefully peeled off the PDMS channel from the mold and punched holes to provide flow inlets, outlets, and pressure sensing ports (diameter $d = 1.07 \pm 0.05$ mm). We also fabricated a thin PDMS film using a Petri dish with the same mixing ratio of 10:1 (w/w) as the top deformable wall. We controlled the thickness of the film by the total mass deposited. After degassing, the film was cured in the oven at $90\si{\degreeCelsius}$ for 30 minutes, with a shorter curing time compared to the thick sidewalls of the channel. Next, we treated the channel and the PDMS thin film with a 4.5 MHz hand-held corona treater (BD-20AC, Electro-Technic Products) for 30 seconds and brought them together into conformal contact for bonding. To confine the deformation solely to the region of interest, we further attached rigid glass slides on top of the thin PDMS film, except for the region of interest to allow a deformable top wall (thickness $t$, see Fig.~\ref{F1}(a)).

For the radially deformable tube, we fabricated a cylindrical tube using a pull-out soft lithography process \cite{raj2018flow}. Specifically, we made a rectangular acrylic cavity to hold a PDMS block. We used a lumbar puncture needle of a nominal diameter of $110~\si{\micro\meter}$ as the replicating mold to fabricate the tube inside the PDMS block. We poured PDMS, with a 10:1 mixing ratio (w/w) of the silicone elastomer base and the curing agent, into the cavity over the needle mold, which was held in position. To provide the pressure sensing ports (diameter $d = 0.43 \pm 0.03$ mm) (see Fig.~\ref{F1}(c)), we used two additional blunt stainless needles as molds. After degassing and curing in the oven at $90\si{\degreeCelsius}$ for 12 hours, we gently removed all the needles and the acrylic cavity from the PDMS block.

In addition, we confirmed all the dimensions of the two fabricated flow geometries by microscope visualization (Table~\ref{T1}). We further measured the Young's modulus of the PDMS, $E$, (both the thin film, \textit{i.e.}, the top wall for the rectangular channel, and the PDMS block from which we fabricated the tube geometry) with a dynamic mechanical analyzer (DMA 850, TA instruments) at $23\si{\degreeCelsius}$. 

\begin{table}
\centering
Channel\hfill\\[1mm]
\begin{tabular}{c@{\quad} c@{\quad} c@{\quad} c@{\quad} c@{\quad} c@{\quad} c@{\quad} c@{\quad}}
    \hline\hline
    $h$ (\si{\milli\meter})  &  $w$ (\si{\milli\meter}) & $\ell$ (\si{\milli\meter}) & $\ell_{\Delta p}$ (\si{\milli\meter}) & $t$ (\si{\milli\meter}) & $d$ (\si{\milli\meter}) & $E$ (\si{\mega\pascal}) & $\nu_\mathrm{s}$ (--)\\
    \hline
    0.25 $\pm$ 0.02  & 5.0 $\pm$ 0.2 & 26.0 $\pm$ 0.3 & 26.0 $\pm$ 0.3 & 0.48 $\pm$ 0.05 & $1.07 \pm 0.05$ & $0.92\pm0.05$ & $0.47\pm0.1$\\
    \hline\hline
\end{tabular}\\
\medskip
Tube\\[1mm]
\begin{tabular}{c@{\quad} c@{\quad} c@{\quad} c@{\quad} c@{\quad} c@{\quad} c@{\quad} c@{\quad} c@{\quad}}
    \hline\hline
    $a_0$ (\si{\micro\meter}) & $w$ (\si{\milli\meter}) & $h$ (\si{\milli\meter}) & $\ell$ (\si{\milli\meter}) & $z_1$ (\si{\milli\meter}) & $\ell_{\Delta p}$ (\si{\milli\meter}) & $d$ (\si{\milli\meter}) & $E$ (\si{\mega\pascal}) & $\nu_\mathrm{s}$ (--)\\
    \hline
    $45\pm1$ & $15\pm{0.1}$ & $7\pm{0.1}$ & $28.7 \pm 0.2$ & $1.4\pm0.1$ & $23.0\pm0.2$ & {$0.43 \pm 0.03$} & $1.58\pm0.08$ & $0.47\pm0.1$\\
    \hline\hline
\end{tabular}   
\caption{Physical parameters and dimensions of the two experimental systems (rectangular channel and axisymmetric tube), for which the pressure drop $\Delta p$ was measured as a function of the flow rate $q$. Here, the Young's modulus for the channel corresponds to that of the top deformable wall.} 
\label{T1}
\end{table}

\subsection{Pressure drop measurement}
\label{sec:pressure_drop_measurement}

To achieve steady flow, we use a syringe pump (11 Pico Plus Elite, Harvard Apparatus) with Teflon tubes. The syringe pump provides a steady fluid flow at a constant volumetric rate $q$ into the inlet of the channel/tube. For the xanthan gum solution, the range of flow rates in experiments was $0.3<q<10~\si{\milli\liter\per\minute}$ (channel) and $5<q<50~\si{\micro\liter\per\minute}$ (tube), while for the glycerin solution, the ranges were $1<q<10~\si{\milli\liter\per\minute}$ (channel) and $1<q<50~\si{\micro\liter\per\minute}$ (tube). The two geometries have different hydraulic resistances, which necessitates different orders of magnitude of the flow rate to achieve pressure drops that lead to similar FSI regimes. The differential pressure drop, $\Delta p$, over the streamwise length, $\ell_{\Delta p}$, was recorded by a pressure sensor (PX26-005DV, OMEGA) with a data acquisition system (DP8PT-006-C24, OMEGA), which continuously acquires the raw data at a rate of 20 Hz. The channel was constructed so that the distance between the pressure ports $\ell_{\Delta p}$ was as close as possible to the streamwise length of the deformable portion $\ell$, \textit{i.e.}, $\ell \approx \ell_{\Delta p}$.  For the tube, the total length $\ell$ of the compliant section (between the rigid inlet and outlet connectors)  is larger than the distance between the pressure ports $\ell_{\Delta p}$, \textit{i.e.},  $\ell > \ell_{\Delta p}$.

\subsection{Preparation and characterization of the shear-thinning and Newtonian fluids}
\label{sec:rheology}

\begin{figure}
    \centering
    \includegraphics[scale=1.5]{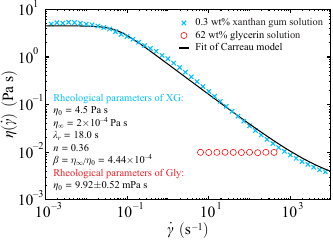}
    \caption{Experimental data for viscosity as a function of shear rate for the xanthan gum (\textcolor{cyan}{$\mathbf{\times}$}) and the glycerin (\textcolor{red}{$\mathbf{\circ}$}) solutions. The solid black curve represents the fit of the xanthan gum solution's rheological data to the Carreau model~\eqref{Carreau model}. The rheological parameters obtained from the fitting are summarized in the figure. 
    Error bars are smaller than the symbols.}
    \label{F2}
\end{figure}

We used 0.3 wt\% xanthan gum (XG, G1253, Sigma Aldrich, molecular weight $\approx 10^6$ g/mol) in deionized (DI) water (pH $\approx 7$) as a representative non-Newtonian shear-thinning fluid with negligible viscoelasticity \cite{raj2018flow, wang2016synergistic,hofmann2003improvement}. Meanwhile, we used a mixture of 62 wt\% of glycerin (Gly) concentration in DI water as a representative viscous Newtonian fluid. We prepared the aqueous solutions by gradually dissolving a known weight of powder or liquid into DI water in a cylindrical beaker. Then, the mixture was continuously stirred for 24 hours until a clear and homogeneous solution was produced. 
We performed rheological measurements of the xanthan gum and glycerin solutions using a controlled-stress rheometer (DHR-3, TA Instruments), which employs a cone-plate geometry (with a diameter of 40 mm and a cone angle of $1^{\circ}$) at a controlled temperature of $25\si{\degreeCelsius}$. The experimentally measured viscosity $\eta(\dot{\gamma})$ as a function of the shear rate $\dot{\gamma}$ for the xanthan gum and glycerin solutions, each averaged across different batches of the solution, is shown in Fig.~\ref{F2}.

To characterize the rheological measurements and the shear-thinning behavior of the xanthan gum solution, we use the Carreau model for the viscosity~\cite{bird1987dynamics1}:
\begin{equation}
    \eta(\dot{\gamma})=\eta_{\infty}+(\eta_{0}-\eta_{\infty})[1+(\lambda_r\dot{\gamma})^{2}]^{(n-1)/2}.
    \label{Carreau model}
\end{equation}
This model captures the three experimentally observed regimes (in Fig.~\ref{F2}) in the viscosity, namely, the plateaus at low and at large shear rates towards $\eta_{0}$ and $\eta_{\infty}$, respectively, and the power-law dependence in-between. The non-integer index $n$ characterizes the degree of shear thinning ($0<n\le 1$) in the power-law regime, and $\lambda_r$ is the inverse of a characteristic shear rate at which shear thinning becomes apparent. The case $n=1$, $\lambda_r=0$, or $\eta_{0}=\eta_{\infty}$ represents the Newtonian fluid with a constant viscosity $\eta_{0}$. The solid black curve in Fig.~\ref{F2} represents the fit of the xanthan gum solution's rheological data to the Carreau viscosity model~\eqref{Carreau model}. 

The three regimes of shear thinning can be quantified by the Carreau number $Cu$ (see, \textit{e.g.}, \cite{shahsavari2015mobility,boyko2021RT,chun2022experimental}), defined as the ratio of the characteristic shear rate of the flow, $q/h_0^2 w$ (channel) or $q/\pi a_0^3$ (tube), to the characteristic shear rate of the fluid, $\lambda_r^{-1}$. Specifically, for the two geometries considered,
\begin{equation}
    Cu_\mathrm{ch} = \frac{\lambda_r q}{h_0^2 w},\qquad
    Cu_\mathrm{tb} = \frac{\lambda_r q}{\pi a_0^3}.
    \label{Cu}
\end{equation}
We note that unlike the rheological parameters $n$, $\lambda_r$, $\eta_0$, $\eta_\infty$, and $\beta=\eta_{\infty}/\eta_{0}$ that remain fixed for the xanthan gum solution, the Carreau number $Cu$ changes through the flow rate $q$, which we vary in our experiments.
Using the values of the geometrical (Table~\ref{T1}) and rheological (Fig.~\ref{F2})
parameters for the range of flow rates achieved in each configuration, we estimated the range of Carreau numbers as: $288<Cu_\mathrm{ch}<9600$ and $5240<Cu_\mathrm{tb}<52397$.

From \eqref{Carreau model}, it is clear that for sufficiently small values of $Cu$, $\eta(\dot{\gamma})\approx\eta_0$, while for sufficiently large values of $Cu$, $\eta(\dot{\gamma})\approx\eta_\infty$. For intermediate values of $Cu$, \eqref{Carreau model} can be approximated as 
\begin{equation}
    \eta(\dot{\gamma}) \approx m|\dot{\gamma}|^{n-1}\qquad\text{with}\qquad m=\eta_{0}\lambda_r^{n-1},
    \label{Power-law model}
\end{equation}
which is the well-known power-law model for viscosity with consistency index $m$. The transition from this power-law regime to the high-$Cu$ plateau can be characterized by a critical Carreau number $Cu_c=\lambda_{r}\dot{\gamma}_{c}=\beta^{1/(n-1)}$ estimated by equating $\eta(\dot{\gamma})$ from \eqref{Power-law model} with $\eta_\infty$. Based on Fig.~\ref{F2}, we have $Cu_c\approx1.72\times10^5$, which is an order of magnitude larger than our estimated values of $Cu_\mathrm{ch}$ and $Cu_\mathrm{tb}$ based on the experimental conditions. Therefore, we conclude that our experiments lie in the power-law regime of shear thinning.

\section{Theory for the flow rate--pressure drop relation incorporating shear thinning}

In this section, we provide the theoretical framework required to rationalize the experimentally obtained $q-\Delta p$ relation and enable a quantitative comparison between the two. We consider low-Reynolds-number flow of incompressible fluids in slender geometries ($a/\ell,h/\ell\ll1$), which allows the use of the lubrication approximation. In this limit, fluid inertia and longitudinal gradients are negligible, and the flow is approximately unidirectional with $\boldsymbol{v} = v_z \boldsymbol{e}_z$ and $p=p(z)$ (see, \textit{e.g.}, \cite{stone2017fundamentals}). Using the generalized Newtonian model to describe shear thinning, the fluid's momentum equations \cite{christov2021soft} reduce to:
\begin{equation}
    0 = - \frac{\mathrm{d}p}{\mathrm{d}z} + \boldsymbol{\nabla}_\perp\boldsymbol{\cdot}[\eta(\dot{\gamma})\boldsymbol{\nabla}_\perp v_z],
    \label{z_lubrication}
\end{equation}
where $\boldsymbol{\nabla}_\perp$ is the gradient in the cross-sectional $(x,y)$ or $(r,\theta)$ coordinates. In~\eqref{z_lubrication}, the viscosity $\eta(\dot{\gamma})$ depends only on the instantaneous shear rate $\dot{\gamma}=\sqrt{2\boldsymbol{E:E}}$, where $\boldsymbol{E}=(\boldsymbol{\nabla}\boldsymbol{v}+(\boldsymbol{\nabla}\boldsymbol{v})^{\mathrm{T}})/2$ is the rate-of-strain tensor. Under the lubrication approximation, $\dot{\gamma} = \partial v_z/\partial y$ (channel) or $\dot{\gamma} = \partial v_z/\partial r$ (tube).

Specifically, the lubrication approximation and \eqref{z_lubrication} apply when the reduced Reynolds number is small. Using a characteristic axial velocity scale $\mathcal{V}_z$, the reduced Reynolds number $Re$ is the ratio of fluid inertia, $\rho\mathcal{V}_z^{2}/\ell$, to viscous stress, $\eta(\dot{\gamma}) \mathcal{V}_z/h_{0}^2$ (channel) or $\eta(\dot{\gamma}) \mathcal{V}_z/a_{0}^2$ (tube). For a shear-thinning fluid in the power-law regime, using \eqref{Power-law model}, we have $Re = Re_{\rm Newt} Cu^{1-n} \ll 1$. For a channel, $Re_{\rm Newt} = \rho \mathcal{V}_z h_0^2/\eta_0 \ell = \rho q h_0 / \eta_{0} w \ell$ and $Cu$ is given in \eqref{Cu}. For a tube,  $Re_{\rm Newt} =\rho \mathcal{V}_z a_0^2/\eta_0\ell = \rho q / \eta_0\pi \ell$ and $Cu$ is given by \eqref{Cu}. Using the values of the geometrical (Table~\ref{T1}) and rheological (Fig.~\ref{F2}) parameters for the range of flow rates (Sec.~\ref{sec:pressure_drop_measurement}) achieved in each configuration, and estimating all working fluids' density as $\rho\approx10^3~\si{\kilo\gram\per\meter\cubed}$, we find that for the xanthan gum solution, $Re_\mathrm{ch}<3\times10^{-2}$ and $Re_\mathrm{tb}<2\times10^{-3}$, while for the glycerin solution, $Re_{\rm Newt, ch}<3\times10^{-2}$ and $Re_{\rm Newt, tb}<10^{-3}$, thus the fluid inertia is indeed negligible.

Certain models for $\eta(\dot{\gamma})$ allow us to integrate \eqref{z_lubrication} to find an analytical expression for $v_z$ \cite{bird1987dynamics1}. From the velocity profile, the volumetric flow rate is obtained as
\begin{equation}
    q = \iint_{\mathcal{A}_\mathrm{deformed}} v_z \,\mathrm{d}A_\perp.
    \label{flow_rate}
\end{equation} 
The cross-sectional domain is either $\mathcal{A}_\mathrm{deformed}=\{(x,y) \,|\, -w/2\le x\le +w/2,0\le y \le h_0+u_y\}$ with $\mathrm{d}A_\perp = \mathrm{d}x\mathrm{d}y$ for the channel ($u_y$ is the vertical displacement of the fluid--solid interface) or $\mathcal{A}_\mathrm{deformed}=\{(r,\theta) \,|\, 0\le r\le a_0+u_r,0\le\theta<2\pi\}$ with $\mathrm{d}A_\perp = r\mathrm{d}r\mathrm{d}\theta$ for the tube ($u_r$ is the radial displacement of the fluid--solid interface). We consider only steady flows, thus $q=\mathrm{const}$. Then, specifically, {we obtain the solution for the axial velocity $v_z$ from the lubrication momentum equation~\eqref{z_lubrication}  using the power-law model~\eqref{Power-law model} for the viscosity, as in \cite{anand2019non,anand2021revisiting}. We obtain the solution for the fluid--solid interface displacement $u_y$ or $u_r$ from the equations of linear elasticity suitable for each geometry (a Reissner--Mindlin plate theory for the channel \cite{anand2019non} and a large block with a circular exclusion under a plane strain configuration for the tube \cite{wang2022flow}). Then,} substituting the latter into  \eqref{flow_rate} yields a nonlinear ODE for the pressure, from which the flow rate--pressure drop relation is determined. {We do not repeat all calculation steps here as they are standard and given in detail in the cited references.}

\subsection{Flow rate--pressure drop relation for a deformable channel}
\label{sec:theory_channel}

For a shear-thinning fluid within the power-law regime, the pressure drop over a streamwise distance $\ell_{\Delta p}$ of a rigid rectangular channel is known \cite{bird1987dynamics1} and can be written as
\begin{equation}
    \Delta p_{\rm rigid, ch}
    = 2(4+2/n)^n \frac{\eta_{0}\ell_{\Delta p}}{\lambda_r h_0} Cu_\mathrm{ch}^{n},
    \label{eq:dp_plaw_ch}
\end{equation}
where $Cu_\mathrm{ch}$ is defined in \eqref{Cu}.
In a slender channel with a deformable top wall (for which the displacement $u_y(x,z)$ obeys Reissner--Mindlin plate theory, such that $\max_{x,z} u_y \ll t < w$ but $t/w\not\to0$), \citet{anand2019non} showed that the pressure satisfies the nonlinear ODE
\begin{equation}
    - \frac{\mathrm{d}p_\mathrm{ch}}{\mathrm{d}z} 
    = \frac{\Delta p_{\rm rigid,ch}}{\ell_{\Delta p}}
    \left\{ 1 + \sum_{k=1}^{\infty} c(k,n) \left[\frac{1}{384\tilde{t}} \frac{\alpha_\mathrm{ch}}{Cu_\mathrm{ch}^n}\frac{p_\mathrm{ch}(z)}{\eta_0\ell/\lambda_r h_0}\right]^{k} 
    \hypgeo{2}{1} \left(\tfrac{1}{2},-k;\tfrac{3}{2}+k;\tilde{t}\right)\right\}^{-n},
    \label{eq:ODE_plaw_ch}
\end{equation}
where $c(k,n) = \sqrt{\pi}\Gamma(3+1/n)/[2\Gamma(3+1/n-k)\Gamma(3/2+k)]$ is related to the generalized binomial coefficient, $\Gamma$ is the Gamma function, and $\hypgeo{2}{1}$ is Gauss' hypergeometric function.  For convenience, we defined  $\tilde{t} = [1 + {8(t/w)^2}/(1-\nu_\mathrm{s})]^{-1}$, where $\nu_s$ is Poisson's ratio. The ODE for a Newtonian fluid is obtained by setting $n=1$, in which case $\lambda_r$ cancels out in \eqref{eq:dp_plaw_ch} and \eqref{eq:ODE_plaw_ch}.

From \eqref{eq:dp_plaw_ch}, the characteristic pressure scale for a shear-thinning fluid flow in the power-law regime is $\mathcal{P}_{\rm flow, ch} = Cu_\mathrm{ch}^n(\eta_0 \ell/\lambda_r h_0)$. We remind the reader that the values of the rheological parameters $\eta_0$, $\lambda_r$, and $n$ are provided in the inset of Fig.~\ref{F2}. Meanwhile, for a plate-like top wall, $\mathcal{P}_{\rm deform, ch}=Bh_0/w^4$ is a characteristic pressure scale for deformation (spanwise bending of the wall), where $B=Et^3/[12(1-\nu_s^2)]$ is the plate's bending rigidity and $E$ is Young's modulus. This ratio of pressure scales is the key dimensionless parameter characterizing the flow-induced deformation \cite{christov2021soft}:
\begin{equation}
    \alpha_\mathrm{ch} 
    = \frac{\mathcal{P}_{\rm flow, ch}}{\mathcal{P}_{\rm deform, ch}} = \left(\frac{\lambda_r q}{h_0^2 w}\right)^{n} \left(\frac{\eta_0 \ell/\lambda_r h_0}{B h_0/w^4}\right)=Cu_\mathrm{ch}^{n}\left(\frac{\eta_0 \ell/\lambda_r h_0}{B h_0/w^4}\right),
\label{eq:alpha_ch}
\end{equation}
which can be termed the \emph{compliance number} for shear-thinning fluids. For $n=1$, \eqref{eq:alpha_ch} reduces to the compliance number $\alpha_{\rm Newt, ch}=\eta_0 \ell q w^3/Bh_0^4$ for a Newtonian fluid in a channel with a compliant top wall~\cite{christov2018flow,martinez2020start}.

For a given $q$, the ODE~\eqref{eq:ODE_plaw_ch} subject to $p_\mathrm{ch}(\ell)=0$ is solved numerically for $p_\mathrm{ch}(z)$ using \texttt{solve\_ivp} from the SciPy stack \cite{SciPy}, and the pressure drop for the deformable channel is calculated as $\Delta p_\mathrm{ch} = p_\mathrm{ch}(0)$. The stiff `LSODA' integration method is used with relative and absolute tolerances of $10^{-12}$, and the series in \eqref{eq:ODE_plaw_ch} is truncated at $50$ terms, having verified that $\Delta p_\mathrm{ch}$ has become independent of the maximum $k$ value.

\subsection{Flow rate--pressure drop relation for a deformable tube}
\label{sec:theory_tube}

For a shear-thinning fluid within the power-law regime, the pressure drop over a streamwise distance $\ell_{\Delta p}$ of an axisymmetric, rigid tube is known \cite{bird1987dynamics1} and can be written as
\begin{equation}
    \Delta p_{\rm rigid, tb} 
    = 2(3+1/n)^n \frac{\eta_0\ell_{\Delta p}}{\lambda_r a_0} Cu_\mathrm{tb}^n,
    \label{eq:dp_plaw_tb}
\end{equation}
where $Cu_\mathrm{tb}$ is defined in \eqref{Cu}.
For a deformable tube extruded from a large block of elastic material, such that $a\ll w$ and $a\ll h$, the displacement solution $u_r(z)$ was found by \citet{wang2022flow} from the equations of linear elasticity under a plane strain configuration. Combining the latter with the results of \citet{anand2021revisiting}, we  obtain a new analytical solution for the pressure profile of a shear-thinning fluid, under the power-law viscosity model~\eqref{Power-law model}, in this geometry:
\begin{equation}   
    p_{\rm tb}(z)  = \frac{ \Delta p_{\rm rigid,tb}}{\ell_{\Delta p}} \frac{\ell}{2(3+1/n)^n \alpha_\mathrm{tb}} \left\{\left[ 1 + (2+3n) 2(3+1/n)^n \alpha_\mathrm{tb} (1-z/\ell)\right]^{1/(2+3n)}-1\right\}.    
    \label{eq:p_plaw_tb_deform}
\end{equation}
Here, $G=E/[2(1+\nu_s)]$ is the elastic shear modulus. In \eqref{eq:p_plaw_tb_deform}, $\ell$ is the \emph{total} length of the deformable section of the tube, subject to the gage pressure condition at its end, $p(\ell)=0$. To compare to the experimental measurements, a partial pressure drop is computed as $\Delta p_\mathrm{tb}= p_\mathrm{tb}(z_1) - p_\mathrm{tb}(z_1 + \ell_{\Delta p})$ from \eqref{eq:p_plaw_tb_deform}.
The solution for a Newtonian fluid is obtained by setting $n=1$, in which case $\lambda_r$ cancels out in \eqref{eq:dp_plaw_tb} and \eqref{eq:p_plaw_tb_deform}. As in Sec.~\ref{sec:theory_channel}, we observe from \eqref{eq:dp_plaw_tb} that $\mathcal{P}_{\rm flow, tb} = Cu_\mathrm{tb}^n(\eta_0 \ell/\lambda_r a_0)$ is the characteristic pressure scale for a shear-thinning fluid flow in the power-law regime, while $\mathcal{P}_{\rm deform, tb}=2G$ is the characteristic pressure scale for deformation (radial expansion of the tube). Thus, the shear-thinning compliance number for our tube configuration is:
\begin{equation}
    \alpha_\mathrm{tb} = \frac{\mathcal{P}_{\rm flow, tb}}{\mathcal{P}_{\rm deform, tb}} 
    = \left(\frac{\lambda_r q}{\pi a_0^3} \right)^{n}\left(\frac{\eta_0 \ell/\lambda_r a_0}{2G}\right)= Cu_\mathrm{tb}^{n}\left(\frac{\eta_0 \ell/\lambda_r a_0}{2G}\right),
    \label{eq:alpha_tb}
\end{equation}
where for a Newtonian fluid with $n=1$, we have $\alpha_{\rm Newt,tb} = \eta_0 \ell q/2\pi G a_0^4$.

\section{Quantitative comparison between theory and experiment and discussion}

\begin{figure}
    \centering
    \includegraphics[scale=1.5]{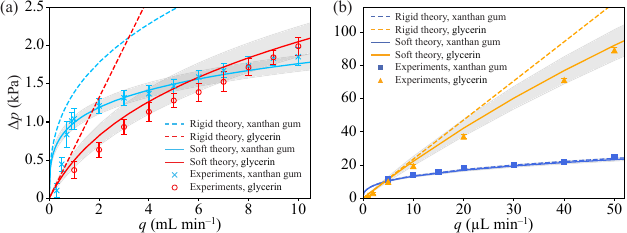}
    \caption{Comparison between our `soft' theory (solid curves), based on \eqref{eq:ODE_plaw_ch} and \eqref{eq:p_plaw_tb_deform},  and the experimental data (symbols) for the flow rate--pressure drop relation in (a) the channel and (b) the tube. The shaded regions indicate the combined uncertainty in (a) $h_0$ or (b) $a_0$ and the viscosity of the glycerin solution. Dashed curves denote the respective rigid-conduit relations (\ref{eq:dp_plaw_ch}) and (\ref{eq:dp_plaw_tb}). Error bars represent the standard deviation based on more than three individual experiments.} 
    \label{F3}
\end{figure}

In Fig.~\ref{F3}, we present a comparison of our theory and the experimental measurements for the flow rate--pressure drop relation of the xanthan gum solution (shear-thinning) and the glycerin solution (Newtonian) in the rectangular channel with a deformable top wall (Fig.~\ref{F3}(a)) and the axisymmetric deformable tube (Fig.~\ref{F3}(b)). Clearly, there is good agreement between our `soft' theoretical predictions and the experimental results, yet theory slightly overpredicts the experimental pressure drop. The major source for this discrepancy is uncertainty in the measurements of $h_0$ and $a_0$ (see Table~\ref{T1}). For the glycerin solution, we also have an uncertainty in the viscosity (see Fig.~\ref{F2}). We, therefore, added shaded regions about the curves in Fig.~\ref{F3} that incorporate the combined uncertainty, obtaining a much better agreement between theory and experiments.

Dashed curves in the figure represent the rigid conduit $q - \Delta p$ relations. For the case of a channel (Fig.~\ref{F3}(a)), we observe ``strong'' fluid--structure interaction, corresponding to large values of compliance numbers in this geometry: $45.9< \alpha_\mathrm{ch} < 162$ (xanthan gum solution) and $12.6 < \alpha_\mathrm{ch} < 126$ (glycerin solution). Therefore, unsurprisingly, the rigid theory prediction~\eqref{eq:dp_plaw_ch} for the channel fails to capture the flow rate--pressure drop relation.
In contrast, for the case of a tube (Fig.~\ref{F3}(b)), the xanthan gum solution exhibits negligible FSI ($3.2\times 10^{-3} < \alpha_\mathrm{tb} < 7.4\times 10^{-3}$), and thus the rigid conduit $q - \Delta p$ relation \eqref{eq:dp_plaw_tb} is valid (the blue solid and dashed curves are almost indistinguishable), while the glycerin solution exhibits weak but measurable fluid--structure interaction  ($3.4 \times 10^{-4} < \alpha_\mathrm{tb} < 1.7\times 10^{-2}$). The apparent difference in the strength of FSI between the channel and tube geometries lies in the fact that the width of the channel has a strong effect (appearing as $w^3$ in $\alpha_\mathrm{ch}$), allowing us to ``increase'' the compliance effect without changing the height. Such a ``compliance tuning'' is not possible for the tube, for which the only cross-sectional dimension is $a_0$, thus making it challenging to measure significant FSI for shear-thinning fluids in this geometry, consistent with previous experiments~\cite{raj2018flow,raj2019biomimetic}. Furthermore, as both shear thinning and compliance lead to a sublinear $q-\Delta p$ relation, an appropriate estimation of $\alpha$, based on \eqref{eq:alpha_ch} and \eqref{eq:alpha_tb} obtained in this work, is required to rationalize which of the two effects is responsible for this nonlinear behavior.

\begin{figure}
    \centering
    \includegraphics[scale=1.5]{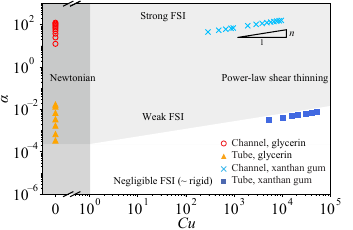}
    \caption{The experimental data from Fig.~\ref{F3} shown on a $Cu-\alpha$ qualitative diagram, in which four regimes of low-Reynolds-number fluid--structure interaction of Newtonian and shear-thinning fluids can be identified (shaded regions). Note the truncated horizontal axis.}
    \label{F4}
\end{figure}

To further delineate the interplay between the rheology of the fluid and the compliance of the deformable conduit, we present in Fig.~\ref{F4} the experimental data on the $Cu-\alpha$ qualitative diagram, showing four distinct regions (physical regimes). The lower half of the diagram corresponds to weak FSI, while the upper half corresponds to strong FSI. Similarly, the left half of the diagram corresponds to negligible shear thinning (Newtonian behavior), while the right half corresponds to significant shear thinning in the power-law regime. The slanted line denoting ``Weak FSI'' is a guide to the eye, as the threshold of ``weak'' is not a strict definition. Our experimental data spans all four regimes and confirms the power-law scaling $\alpha \sim Cu^n$ from \eqref{eq:alpha_ch} and \eqref{eq:alpha_tb}, for the xanthan gum solution. Nevertheless, we do not observe the transition from the low-$Cu$ to the power-law behavior and the transition from the power-law to the high-$Cu$ behavior, as our $q-\Delta p$ data correspond to intermediate values of the Carreau number (see Sec.~\ref{sec:rheology}). For example, in the case of the channel, achieving the low-$Cu$ regime with a non-negligible FSI would require providing very small flow rates while increasing the length of the channel. On the other hand, the high-$Cu$ regime would require providing large flow rates, resulting in large shear rates that may lead to inertial effects and viscous heating~\cite{pipe2008high}.

\section{Concluding remarks}

In this work, we analyzed the interplay between shear-thinning rheology and wall compliance and provided the first quantitative comparison between theory and experiments on low-Reynolds-number shear-thinning fluid flows in two canonical deformable geometries. We showed good agreement for the steady flow rate--pressure drop relation given by the theory, which has no fitting parameters, and experiments, which we ensured are in the regime of significant flow-induced deformation.
Given common experiment/theory discrepancies for complex fluids (see, \textit{e.g.}, \cite{sanchez2022understanding}), we believe that such a quantitative comparison is of fundamental importance in non-Newtonian fluid mechanics since it, and not a comparison of theory with simulations, serves as a real validation of the adequacy of the constitutive model used.

Having experimentally demonstrated the quantitative predictive power of our theory of the flow rate--pressure drop relation of Newtonian and shear-thinning fluids in two canonical deformable geometries with experiments, as a future research direction, it is interesting to perform experiments in the rectangular geometry used herein with viscoelastic fluids. Similar to previous experimental studies in rigid non-uniform geometries~\cite{rothstein1999extensional,rothstein2001axisymmetric,campo2011flow,ober2013microfluidic}, the experimental set-up for viscoelastic fluids will consist of two long straight channels connected to the deformable region upstream and downstream to eliminate the entrance and exit effects. Furthermore, since the flow of viscoelastic fluids may become unstable above a certain flow rate due to the fluid’s complex rheology~\cite{larson1992instabilities,shaqfeh1996purely,steinberg2021elastic,datta2021perspectives}, the experiments will require extra care in measuring the flow rate--pressure drop relation in steady and stable flows. Such experiments will enable a quantitative comparison with a recent theory of \citet{boyko2023non}.

\begin{acknowledgments}
S.C.\ and J.F.\ acknowledge the use of facilities and instrumentation at the Materials Research Laboratory Central Research Facilities, University of Illinois, for the rheological experiments. J.F.\ gratefully acknowledges partial support from the American Chemical Society Petroleum Research Fund Grant No.\ 61574-DNI9. E.B.\ gratefully acknowledges support from the Israel Science Foundation (Grant No.\ 1942/23).
\end{acknowledgments}

\bibliography{literature}

\end{document}